\documentclass[fleqn,twoside,twocolumn,nofootinbib,showkeys]{revtex4} 
\usepackage[nocpr]{ujp} 

\begin{document}
\title[Variational Method for the Calculation of Critical Distance]
{VARIATIONAL METHOD\\ FOR THE CALCULATION OF CRITICAL DISTANCE\\
BETWEEN TWO COULOMB CENTERS IN GRAPHENE}
\author{O.O.~Sobol}
\affiliation{Taras Shevchenko National University of Kyiv}
\address{2, Prosp. Academician Glushkov, Kyiv 03022, Ukraine}
\email{sobololeks@ukr.net}

\udk{538.915} \pacs{81.05.ue, 73.22.Pr} \razd{\secviii}

\autorcol{O.O.\hspace*{0.7mm}Sobol}

\setcounter{page}{531}%

\begin{abstract}
The supercritical instability in a system of two identical charged
impurities in gapped graphene described in the continuous limit by
the two-dimensional Dirac equation has been studied.\,\,The case
where the charge of each impurity is subcritical, but their sum
exceeds the critical value calculated in the version with a single
Coulomb center, is considered.\,\,Using the developed variational
method, the dependence of the critical distance $R_{\rm cr}$ between
the impurities on their total charge is calculated.\,\,The $R_{\rm
cr}$-value is found to grow as the total impurity charge increases
and the {quasiparticle band gap decreases}.\,\,The results of
calculations are compared with those obtained in earlier researches.
\end{abstract}
 \keywords{graphene, supercritical instability, critical distance,
Kantorovich variational method.}

\maketitle

\section{Introduction}

The gapless linear spectrum of graphene was determined rather long
ago \cite{Wallace}, while constructing the band model of graphite
provided that the interaction between the planes of carbon atoms is
neglected.\,\,Nevertheless, graphene itself was obtained for the
first time only in 2004 \cite{Geim}, which launched its intense
theoretical and experimental study.\,\,Graphene permanently attracts
attention of scientists throughout the world owing to its unique
properties.\,\,In particular, it is one of the first two-dimensional
crystals and possesses the ultrahigh mechanical strength and the
charge carrier mobility, which determines the prospects of its
application in novel nanoelectronics.\,\,The physics of graphene is
also of interest from the viewpoint of fundamental scientific
researches, because it turned out to have a deep relation to quantum
electrodynamics (QED) and other quantum field theories.

Graphene has a honeycomb crystalline lattice.\,\,In the
tight-binding approximation, the spectrum of low-energy
quasiparticle excitations is linear, and the latter are described by
a $2+1$-dimensional massless Dirac equation
\cite{Semenoff,CastroNeto,Gusynin,Abergel}.\,\,Hence, in the
continuous limit, we obtain an effective quantum field theory with
$2+1$-dimensional Dirac fermions that interact with the ordinary
three-dimensional Coulomb potential, $\sim 1/r$.\,\,Those
circumstances testify to a capability of the solid-state
implementation of experiments aimed at detecting such QED phenomena
as the Klein paradox, Schwinger effect (the creation of pairs in a
strong external electric field), supercritical atomic collapse, and
others, which have not yet been observed in the Nature.\,\,The Klein
tunneling was really revealed in graphene \cite{Kim}, and the
supercritical atomic collapse of charged impurities was recently
observed experimentally \cite{Wang}.

The solution of the relativistic Kepler problem with regard for the
finite size of an atomic nucleus \cite{finite-size} showed that, if
the so-called \textit{critical} charge of a nucleus,
$Z_{\mathrm{cr}}\approx170$, is exceeded, the energy levels of bound
states dives into the lower continuum, and the escape of positrons
is observed \cite{Greiner,Zeldovich}.\,\,However, there are no
nuclei with this charge, and, hence, the effect has not been
observed.\,\,Somewhat later, there emerged an idea concerning
head-on (or almost head-on) collisions between the nuclei of heavy
atoms, e.g., uraniums
\cite{Gershtein,Rafelski,Muller,Zeldovich}.\,\,In this case, the
total charge of the nuclei exceeds the critical value, and there
exists a distance between the nuclei, at which the lowest bound
state dives into the lower continuum.\,\,This distance is also
called \textit{critical}.\,\,Unfortunately, in the relativistic
problem of two centers, the variables cannot be separated in any
coordinate system, so that an analytical solution cannot be derived
\cite{Greiner,Zeldovich}.\,\,However, the calculations with the help
of approximate quantum-mechanical methods, in particular, the
variational one \cite{Popov}, were carried out, and the dependences
of the critical distance between the nuclei on the total system
charge were plotted.

\begin{figure}
\vskip1mm
  \includegraphics[width=7cm]{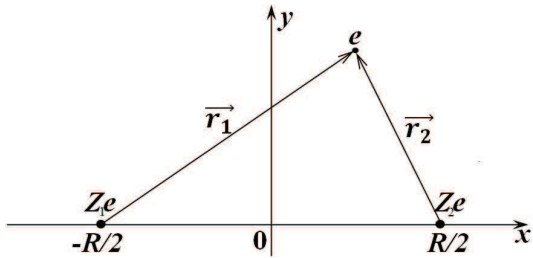}
\vskip-3mm  \caption{Electron in the field of two Coulomb
impurities}\vspace*{2mm}
  \label{fig1}
\end{figure}

In the physics of graphene, the Fermi velocity,
\mbox{$v_{\mathrm{F}}\approx 10^{6}~\mathrm{m/s}$}, appears instead
of the light ve\-lo\-ci\-ty $c$.\,\,As a result, the
\textquotedblleft fine structure constant\textquotedblright\ for
gra\-phe\-ne is $\alpha=\frac{e^{2}}{\hbar
v_{\mathrm{F}}}\approx2.19$.\,\,Ho\-we\-ver, gra\-phe\-ne is
actually located on a dielectric substrate.\,\,Therefore, the
interaction with an impurity with charge $Ze$ is described by the
constant $\beta =Z\alpha/\kappa$, where $\kappa$ is the dielectric
permittivity of a sub\-stra\-te.\,\,The problem of supercritical
instability of a single charged impurity in gapless gra\-phe\-ne was
studied in detail in works \cite{Shytov,Pereira}.\,\,In the case of
gra\-phe\-ne with the gap $2\Delta$ and the regularized Coulomb
potential $V(r)=-\frac{Ze^{2}}{\kappa r}\theta(r-R)-$
$-\frac{Ze^{2}}{\kappa R}\theta(R-r)$, the transition to the
supercritical mode was found to occur at $\beta
_{\mathrm{c}}=1/2+\pi^{2}/\log^{2}(c\Delta R/\hbar v_{\mathrm{F}})$,
where $c\approx0.21$.

Although the supercritical instability manifests itself in graphene
already at impurity charges $Z\gtrsim1$, a possibility to detect it
is illusory because of experimental difficulties associated with the
production of impurities, whose charge is larger than 1.\,\,The
solution of this task consists in arranging a few charged impurities
on a small section of graphene.\,\,Just this approach was
implemented by a group of scientists from the University of
California \cite{Wang}, while observing the supercritical
instability in graphene (ionized dimers of calcium were used as
impurities).\,\,Hence, despite that the coupling constant in
graphene is larger by a factor of 300, the necessity to consider a
few, rather than one, impurities located closely to one another
arises in this case as well, the configuration with two Coulomb
centers being the simplest variant.\,\,In work \cite{two-centers},
the critical distance between impurities was calculated, and the
position and the width of a resonance that arises in the lower
continuum in the supercritical mode were determined in the first
approximation with the use of variational technique.

This work aims at obtaining the next (second) approximation in the
framework of the variational method.\,\,The article structure is as
follows.\,\,In Section~\ref{probl}, the problem is formulated, and
the asymptotics for the wave function of a state that dives into the
lower continuum are determined.\,\,The variational method to find
the critical distance between impurities is described in
Section~\ref{var}.\,\,In Section~\ref{concl}, the results obtained
are discussed, and some conclusions are made.

\section{Statement of the Problem}

\label{probl}

In the tight-binding approximation, the band spectrum of graphene
has a valence band and a conduction band that touch each other at
two nonequivalent points of the reciprocal crystalline
lattice.\,\,These are the so-called \textit{Dirac points~}${\bf
K}_{\pm}$.\,\,The spectrum of low-energy excitations is
linear.\,\,The effective Hamiltonian that describes electron
quasiparticle excitations in vicinities of the Dirac points has the
form of a $2+1$-dimensional Dirac Hamiltonian.\,\,In the case where
there is a quasiparticle gap between the valence and conduction
bands, the Hamiltonian also includes the mass term
\begin{equation}
H=v_{\mathrm{F}}\boldsymbol{\tau}\boldsymbol{p}+\xi\Delta\tau_{z}+V(r),
\end{equation}
where $\boldsymbol{p}$ is the canonical momentum, $\tau_{i}$ are the
Pau\-li matrices, and $\Delta$ is the quasiparticle gap
half\-width.\,\,This Hamiltonian operates in the space of
two-component spinors $\Psi_{\xi s}$ distinguished by the valley
($\xi=\pm$) and spin ($s=\pm$) subscripts.\,\,It is general\-ly
adopted that $\Psi_{+s}^{T}=(\psi_{A},\psi_{B})_{K_{+}s}$ and
$\Psi_{-s}^{T}=$ $=(\psi _{B},\psi_{A})_{K_{-}s}$, where $A$ and $B$
are the corresponding sublattices of the hexagonal lattice in
graphene. The interaction potential reads
\begin{equation}
V\left(\mathbf{r}\right)=-\frac{e^2}{\kappa}\left(\!\frac{Z_1}{r_1}+\frac{Z_2}{r_2}\!\right)\!,
\end{equation}
where $\kappa$ is the dielectric permittivity of the substrate,
$Z_{1,2}$ are the impurity charges, and
$r_{1,2}=|\mathbf{r}\pm\mathbf{R}/2|$ are the distances reckoned
from the Coulomb impurities to the electron (see Fig.~1).\,\,The
potential does not depend on the spin, so that we will omit the spin
subscript below.\,\,We also suppose that the electron is located
near the Dirac point $\boldsymbol{K}_{+}$.\,\,If the electron is
located near the point $\boldsymbol{K}_{\_}$, $\Delta$ has to be
substituted by $-\Delta$ everywhere.\,\,In recent experiments
\cite{Wang}, identical impurities were considered, and we also put
$Z_{1}=$ $=Z_{2}=Z$ in this work.

Hence, the Dirac equation for an electron in the potential of two charged
impurities looks like
\begin{equation}
\left(v_{\rm F}\tau_xp_x+v_{\rm
F}\tau_yp_y+\Delta\tau_z+V\left(\mathbf{r}\right)\right)\Psi(\mathbf{r})
=E\Psi(\mathbf{r}). \label{Dirac-equation}
\end{equation}
For the two-component spinor $\Psi(\mathbf{r})=(\phi,\chi)^{T}$, it
can be rewritten in the form
\begin{equation}
\left\{\!\!\!
\begin{array}{l}
\displaystyle\left(E-V-\Delta\right)\phi+iv_{\rm
F}\left(\!\frac{\partial}{\partial x}-i\frac{\partial}{\partial
y}\!\right)\chi=0,\\[3mm]
\displaystyle\left(E-V+\Delta\right)\chi+iv_{\rm
F}\left(\!\frac{\partial}{\partial x}+i\frac{\partial}{\partial
y}\!\right)\phi=0.
\end{array}
\right. \label{system}
\end{equation}
Expressing the component $\chi$ from the second equation of
system (\ref{system}) and substituting it to the first one, we obtain the
following second-order differential equation for the spinor component
$\phi$:
\[
(\partial^2_x+\partial^2_y)\phi+\frac{\frac{\partial V}{\partial
x}-i\frac{\partial V}{\partial y}}{E-V+\Delta}
\left(\!\frac{\partial\phi}{\partial
x}+i\frac{\partial\phi}{\partial y}\!\right)+
\]\vspace*{-5mm}
\begin{equation}\label{upper-component}
+\,v^{-2}_{\rm F}\left((E-V)^2-\Delta^2\right)\phi=0.
\end{equation}
The supercritical instability takes place when the lowest bound
state dives into the lower continuum, i.e.\,\,when $E=-\Delta$
\cite{Zeldovich,Greiner}.\,\,In the further calculations, only this
energy value will be considered.

\subsection{Asymptotic behavior of the wave function}

Let us analyze the asymptotic behavior of the wave function at large
distances, $r\rightarrow\infty$.\,\,In this limit, the potential
looks like
\begin{equation}
V\left(\mathbf{r}\right)=-\zeta v_{\rm
F}\left(\!\frac{1}{r}+\frac{R^2}{4r^3}
P_2(\cos\varphi)+O\left(\!\frac{1}{r^5}\!\right)\!\right)\!,
\label{potential-asymptotic}
\end{equation}
where $\zeta=2Z\alpha/\kappa$, and $P_{2}(x)$ is the Legendre
polynomial of the second order.\,\,Substituting potential
(\ref{potential-asymptotic}) into Eq.~(\ref{upper-component}) and
preserving only the most contributing terms, we obtain
\begin{equation}
\label{e2}
\phi''+\frac{2}{r}\phi'+\left(\!\frac{\zeta^2}{r^2}-\frac{2m\zeta}{r}\!\right)\phi=0,
\end{equation}
where $m=\Delta/v_{\mathrm{F}}$ is the inverse Compton wavelength of
quasiparticles.\,\,The solution of this equation vanishing at
infinity is described by the Macdonald function as follows:
\begin{equation}
\phi(r)=Cr^{-1/2}K_{i\gamma}(\sqrt{8m\zeta r}),\quad\gamma=\sqrt
{4\zeta^{2}-1}. \label{phi-r-to-infinity}%
\end{equation}
Then, with regard for the asymptotic behavior of the Macdonald function
\cite{Bateman}, we have
\begin{equation}
\phi_{\mathrm{asym}}(r)=\tilde{C}r^{-3/4}\exp(-\sqrt{8m\zeta
r}),\quad
r\rightarrow\infty. \label{asymptotic}%
\end{equation}

To study the asymptotic behavior of the solution in a vicinity of either of
the impurities, it is convenient to change to the elliptic coordinate system
($\xi$, $\eta$), where
\begin{equation}
\xi\equiv\frac{r_{1}+r_{2}}{R},~~~~~\eta\equiv\frac{r_{1}-r_{2}}{R}.
\end{equation}
The new coordinates can be varied within the in\-tervals%
\[
1\leq\xi<\infty,~~~~~ -1\leq\eta\leq1,
\]
and the impurities are located at the points $(\xi=1,$ $\eta=\pm1)$.

In the elliptic coordinates, the interaction potential looks like
\begin{equation}
V\left(  {\bf r}\right)  =-\frac{2\zeta
v_{\mathrm{F}}\xi}{R(\xi^{2}-\eta ^{2})}.
\end{equation}
In a vicinity of either of the impurities, the quantity $\xi^{2}-\eta^{2}$
is small. In the elliptic coordinates, Eq.~(\ref{upper-component}) reads
\[
 \frac{4}{R^2\left(\xi ^2-\eta ^2\right)} \biggl[\sqrt{\xi
^2-1}\frac{\partial}{\partial\xi}\left(\!\sqrt{\xi
^2-1}\frac{\partial} {\partial\xi}\!\right)+
\]\vspace*{-5mm}
\[
+\,\sqrt{1-\eta ^2}\frac{\partial}{\partial\eta}\left(\sqrt{1-\eta
^2}\frac{\partial} {\partial\eta}\right)\!\biggr]\phi\,+
\]\vspace*{-5mm}
\[
+\,\frac{4}{R^2\xi(\xi^2-\eta^2)^3}\biggl[\xi^4\eta+3\xi^2\eta^3-3\xi^2\eta-\eta^3\,-
\]\vspace*{-5mm}
\[
-\,i\sqrt{(\xi^2-1)(1-\eta^2)}(\xi^3+3\xi\eta^2)\!\biggr]\times
\]\vspace*{-5mm}
\[
\times\biggl[\!\left(\eta(\xi^2-1)+i\xi\sqrt{(\xi^2-1)(1-\eta^2)}\right)
\frac{\partial \phi}{\partial\xi}\,+
\]\vspace*{-5mm}
\[
+\left(\xi(1-\eta^2)-i\eta\sqrt{(\xi^2-1)(1-\eta^2)}\right)\frac{\partial
\phi}{\partial\eta}\biggr]+
\]\vspace*{-5mm}
\begin{equation}\label{e3}
+\left(\!\frac{4\zeta^2\xi^2}{R^2(\xi^2-\eta^2)^2}-\frac{4\zeta
m\xi}{R(\xi^2-\eta^2)}\!\right)\phi=0.
\end{equation}
We seek a solution in the form $\phi(\xi,\eta)=\phi(\mu)$, where
$\mu=\xi^2-\eta^2=4r_1r_2/R^2$.\,\,After the substitution in
Eq.\,\,(\ref{e3}), we retain only the most weighty terms and obtain
the equation
\begin{equation}
\frac{d^2\phi}{d\mu^2}+\frac{2}{\mu}\frac{d\phi}{d\mu}+\frac{\zeta^2}{4\mu^2}\phi=0,
\end{equation}
whose solution regular as $\mu \to 0$ is
\begin{equation}
\phi_{\mathrm{imp}}(\mu)=C_{2}\mu^{-\sigma/2},\quad\sigma=1-\sqrt{1-\zeta^{2}%
}. \label{phi-mu-to-zero}%
\end{equation}

Taking into account that $\mu\simeq4r^{2}/R^{2}$ as $r\rightarrow\infty$,
solution (\ref{phi-r-to-infinity}) can be rewritten in the form
\begin{equation}
\phi(\mu)=C_{1}\mu^{-1/4}K_{i\gamma}(2\sqrt{m\zeta R}\mu^{1/4}).
\label{phi-mu-to-infinity}%
\end{equation}
Matching solutions (\ref{phi-mu-to-zero}) and (\ref{phi-mu-to-infinity})
across the point $\mu=1$, we obtain an implicit equation for the dependence of
the critical distance $R_{\mathrm{cr}}$ on $\zeta$, namely, the transcendental
equation
\begin{equation}
2\sqrt{1-\zeta^{2}}-1=2\sqrt{m\zeta R}\frac{K_{i\gamma}^{\prime}(2\sqrt{m\zeta
R})}{K_{i\gamma}(2\sqrt{m\zeta R})}. \label{trancend-eq}%
\end{equation}
Its numerical solution is shown by the dash-dotted curve in
Fig.~2.\,\,In the next section, this dependence will be calculated
more accurately with the help of the variational method.

\section{Variational Method}

\label{var}

As was indicated above, the relativistic Dirac equation, if being
applied to the problem of two Coulomb centers, does not allow the
variables to be separated in any orthogonal coordinate system; that
is why it is impossible to obtain its solution in the analytical
form.\,\,While constructing an approximate solution, let us take
advantage of the variational method, as was done in the case of QED
\cite{Popov}.\,\,In work \cite{variational}, it was indicated that
the highest accuracy within the variational method is achieved when
the corresponding trial functions satisfy the asymptotics of the
exact solution in vicinities of charged impurities and at infinity.
Those asymptotics were obtained in the previous section.

In order to formulate the variational problem, it is enough to mark that
the differential equation (\ref{upper-component}) can be derived, while analyzing
the following functional with respect to the \textrm{extremum} value:
\[
S[\phi]=\int\biggl(\!(E-V+\Delta)^{-1}\left|\frac{\partial\phi}{\partial
x} +i\frac{\partial\phi}{\partial y}\right|^2-
\]
\begin{equation}
-\,v^{-2}_{\rm F}(E-V-\Delta)|\phi|^2\!\biggr)dxdy,
\label{functional}
\end{equation}
and providing the norm preservation condition,
\[
N=\int\Psi^*\Psi dxdy=\int \Biggl[v_{\rm F}^{-2}|\phi|^2\,+
\]\vspace*{-7mm}
\begin{equation}
+\,(E-V+\Delta)^{-2}\left|\frac{\partial\phi}{\partial
x}+i\frac{\partial\phi}{\partial y}\right|^2\Biggr]dxdy.
\end{equation}
Let us designate the new field as $\psi=W^{-1/2}\phi$, where
$W=E-V+\Delta$.\,\,Then the functional $S[\phi]$ can be expressed in
the form typical of non-relativistic quantum mechanics,
\[
S[\psi]=\int\biggl[|\bm\nabla\psi|^2+i\left(\!\frac{\bm\nabla V}{2W}
\times\bm\nabla\psi^*\!\right)\psi\,-
\]\vspace*{-5mm}
\begin{equation}
-\,i\psi^* \left(\!\frac{\bm\nabla
V}{2W}\times\bm\nabla\psi\!\right)+2(U-\epsilon)|\psi|^2\biggr]dxdy,
\label{functionalS-modified}
\end{equation}
where $\mathbf{a}\times\mathbf{b}=\epsilon_{ij}a_{i}b_{j}$, $\epsilon
=(E^{2}-\Delta^{2})/2v_{\mathrm{F}}^{2}$ is the effective energy, and the
effective potential $U$ looks like
\begin{equation}
U=\frac{EV}{v_{\mathrm{F}}^{2}}-\frac{V^{2}}{2v_{\mathrm{F}}^{2}}%
+\frac{\triangle V}{4W}+\frac{3}{8}\frac{(\bm\nabla V)^{2}}{W^{2}}.
\end{equation}
The second and third terms in functional
(\ref{functionalS-modified}) describe the pseudospin-orbit
coupling.\,\,The norm reads
\[
N=\int\Biggl[|\bm\nabla\psi|^2\,+
\]\vspace*{-5mm}
\[
+\,i\left(\!\frac{\bm\nabla
V}{2W}\times\bm\nabla\psi^*\!\right)\psi-i\psi^*
\left(\!\frac{\bm\nabla V}{2W}\times\bm\nabla\psi\!\right)+
\]\vspace*{-5mm}
\begin{equation}
+\left(\!\frac{W^2}{v_{\rm F}^2}+\frac{\triangle
V}{2W}+\frac{5(\bm\nabla V)^2}{4W^2}\!\right)
|\psi|^2\Biggr]W^{-1}dxdy,
\end{equation}
being important if proper boundary conditions are selected.\,\,In
what follows, we are interested in the case where the lowest bound
state dives into the lower continuum.\,\,Therefore, we put
$E=-\Delta$ ($\epsilon=0$) and $W=-V$.\,\,As a result, the
functional $S[\psi]$ becomes simpler.

\begin{figure*}
\vskip1mm
   \includegraphics[width=12.5cm]{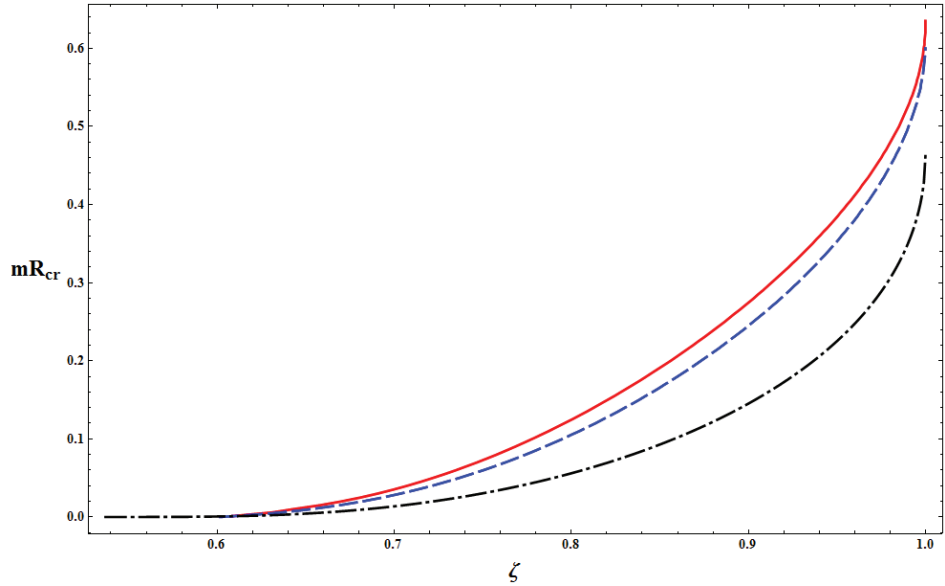}
   \vskip-3mm
\parbox{12.5cm}{\caption{Dependences $mR_{\mathrm{cr}}(\zeta)$ obtained as
numerical solutions of Eqs.~(\ref{trancend-eq}) (dash-dotted curve)
and (\ref{condition}) (solid curve).\,\,For comparison, the results
of calculations obtained in the first approximation of the
variational method and taken from work \cite{two-centers} are
exhibited by the dashed curve } \label{fig2}}
\end{figure*}

In Section \ref{probl}, the exact solution of the problem was found
to depend only on the single variable $\mu$ in both asymptotic cases
(\ref{phi-mu-to-zero}) and (\ref{phi-mu-to-infinity}).\,\,Therefore,
we use the Kantorovich variational method (for more details, see
Section~3 in review \cite{Popov-review}) and select the sums
\begin{equation}
\psi=\sum\limits_{k=1}^{N}\psi_{k}(\mu)\nu^{k-1} \label{Kantorovich-method}%
\end{equation}
as trial functions.\,\,Here, $\psi_{k}(\mu)$ compose a set of sought
functions, and $\nu=\nu(\xi,\eta)$ is a fixed function of the
coordinates, which is independent of $\mu$.\,\,In work \cite{Popov},
two variants of the function $\nu$ were considered.\,\,The results
obtained turned out close enough.\,\,Therefore, in
this work, we select the variant $\nu=\eta^{2}/(\xi^{2}-\eta^{2})=$ $=(r_{1}%
-r_{2})^{2}/4r_{1}r_{2}$.

Substituting expression~(\ref{Kantorovich-method}) into functional
(\ref{functionalS-modified}) and integrating over $\nu$, we obtain
\[
S_N(\psi)=4\sum\limits_{k,l=1}^{N} \int\limits_{0}^{\infty}d\mu
\biggl(\!P_{kl}\psi_k'{\psi^*_l}' +Q_{kl}\psi_k\psi^*_l+
\]\vspace*{-5mm}
\begin{equation}
+\,R_{kl}\psi_k'\psi^*_l+R_{kl}^{\dagger}{\psi_k}{\psi^*_l}'\!\biggr)\!,
\label{functional2}
\end{equation}
where $\mathbf{\hat{P}}$, $\mathbf{\hat{Q}}$, and $\mathbf{\hat{R}}$
are $N\times N$-matrices depending on $\mu$.\,\,The corresponding
expressions for them are given in Appendix~A (see
Eqs.~(\ref{P})--(\ref{R})).

Functional (\ref{functional2}) reaches its minimum on the solutions of
the Lagrange--Euler equations
\begin{equation}
\frac{d}{d\mu}\left(\!P_{kl}\frac{d\psi_k}{d\mu}+R_{kl}^{\dagger}\psi_k\!\right)-Q_{kl}\psi_k-R_{kl}
\frac{d\psi_k}{d\mu}=0, \label{set-of-equations}
\end{equation}
where $l=\overline{1,N}$.\,\,The boundary conditions are so selected
that the norm remains finite, and the trial functions satisfy the
exact solution asymp\-to\-tics.\,\,The Lag\-ran\-ge--Euler equations
and the corresponding boundary conditions constitute a boundary
problem to be solved.

In work \cite{two-centers}, the first approximation of the
Kantorovich variational method with $N=1$ was used. The shooting
method was used to calculate the dependence of the critical distance
between two impurities on their total charge,
$R_{\mathrm{cr}}(\zeta)$.\,\,The result obtained is shown in Fig.~2
by the dashed curve for comparison.

A higher accuracy can be reached by applying the Kantorovich
variational method with $N>1$. Ho\-we\-ver, in this case, there are
some differences from the case $N=1$ considered in work
\cite{two-centers}.\,\,First of all, the shooting method is no more
applicable, because the system of differential equations rather than
a single equation is dealt with.\,\,It was noticed in work
\cite{Marinov}, that, in order to simplify numerical calculations,
the system of differential equations of the second order has to be
reduced to a matrix differential equation of the first order with
the use of the substitution\vspace*{-1mm}
\begin{equation}
\psi_{i}^{\prime}(\mu)=\sum\limits_{j=1}^{N}Y_{ij}(\mu)\psi_{j}(\mu).
\end{equation}
In this case, we obtain the following nonlinear matrix Riccati equation for
the matrix $\mathbf{\hat{Y}}$:
\begin{equation}
\mathbf{\hat{Y}}^{\prime}=\mathbf{\hat{A}}-\mathbf{\hat{B}}\mathbf{\hat{Y}%
}-\mathbf{\hat{Y}}^{2}.
\end{equation}
Here, the matrix coefficients $\mathbf{\hat{A}}$ and $\mathbf{\hat{B}}$ look
like
\begin{equation}
\mathbf{\hat{A}}=\mathbf{\hat{P}}^{-1}(\mathbf{\hat{Q}}-\mathbf{\hat{R}%
}^{\prime}),
\end{equation}\vspace*{-7mm}
\begin{equation}
\mathbf{\hat{B}}=\mathbf{\hat{P}}^{-1}(\mathbf{\hat{R}}-\mathbf{\hat{R}}%
^{T}+\mathbf{\hat{P}}^{\prime}).
\end{equation}

The matrices $\mathbf{\hat{P}}$, $\mathbf{\hat{Q}}$, and
$\mathbf{\hat{R}}$ are singular at zero and infinity.\,\,This is a
reflection of the fact that the different functions $\psi_{k}(\mu)$,
as well as different elements of the matrix $\mathbf{\hat{Y}}$, are
characterized by different growth/fall degrees in vicinities of
special points.\,\,To overcome this inconvenience, let us subject
the trial functions to a linear transformation,
\begin{equation}
\psi_{k}(\mu)=S_{kl}(\mu)\tilde{\psi}_{l}(\mu),\quad\det\mathbf{\hat{S}}%
(\mu)\neq0. \label{transf}%
\end{equation}
The new matrix is designated as
$\mathbf{\hat{Y}}_{\mathbf{S}}$,\vspace*{-1mm}
\begin{equation}
\tilde{\psi}_{k}^{\prime}(\mu)=\left(  Y_{\mathbf{S}}\right)  _{kl}(\mu
)\tilde{\psi}_{l}(\mu).
\end{equation}
It is related to the matrix $\mathbf{\hat{Y}}$ by the
formula\vspace*{-1mm}
\begin{equation}
\mathbf{\hat{Y}}_{\mathbf{S}}=\mathbf{\hat{S}}^{-1}\mathbf{\hat{Y}%
}\mathbf{\hat{S}}-\mathbf{\hat{S}}^{-1}\mathbf{\hat{S}}^{\prime}.
\end{equation}
The matrix $\mathbf{\hat{Y}}_{\mathbf{S}}$ also satisfies the
Riccati equation,\vspace*{-1mm}
\begin{equation}
\mathbf{\hat{Y}}_{\mathbf{S}}^{\prime}=\mathbf{\hat{A}}_{\mathbf{S}%
}-\mathbf{\hat{B}}_{\mathbf{S}}\mathbf{\hat{Y}}_{\mathbf{S}}-\mathbf{\hat{Y}%
}_{\mathbf{S}}^{2}. \label{Riccati-eq}%
\end{equation}

The laws governing the change of coefficients in
Eqs.~(\ref{Riccati-eq}) and (\ref{set-of-equations}) at
transformation (\ref{transf}) are given in Appen\-dix~B.\,\,They
give rise to the criterion of choice of a matrix $\mathbf{\hat{S}}$;
namely, the matrix $\mathbf{\hat{P}}_{\mathbf{S}}$ must be
nondegenerate after the transformation, because the corresponding
inverse matrix must exist as well.

In vicinities of each special points ($\mu=0$ and
\mbox{$\mu\rightarrow\infty$}), the corresponding specific
transformation matrix $\mathbf{\hat{S}}^{0/\infty}$
must be selected, and the transformed matrix coefficients $\mathbf{\hat{A}%
}_{\mathbf{S}}$ and $\mathbf{\hat{B}}_{\mathbf{S}}$ must be expanded in
series in the variable $\mu$.\,\,Accordingly, the sought matrix $\mathbf{\hat{Y}%
}_{\mathbf{S}}$ should be of the form
\begin{equation}
\begin{array}{l}
\displaystyle \mathbf{\hat{Y}}_{\mathbf{S}}^{(0)}=\mu^{\lambda_0}
\left(\!\mathbf{\hat{Y}}_{1}^{(0)}+ \mu
\mathbf{\hat{Y}}_{2}^{(0)}+\mu^2 \mathbf{\hat{Y}}_{3}^{(0)}+...
\!\right)\!,\  \mu\to 0,\\[3mm]
\displaystyle
\mathbf{\hat{Y}}_{\mathbf{S}}^{(\infty)}\!=\!\mu^{\lambda_{\infty}}
\left(\!\mathbf{\hat{Y}}_{1}^{(\infty)}\!+\!\frac{1}{\mu}
\mathbf{\hat{Y}}_{2}^{(\infty)}\!+\!\frac{1}{\mu^2}
\mathbf{\hat{Y}}_{3}^{(\infty)}\!+... \!\right)\!,\\[3mm]
\displaystyle \mu\to \infty.
\end{array}\!\!\!\!\!\!\!\!\!\!
\end{equation}
Let us substitute those formulas in Eq.\,(\ref{Riccati-eq}) and
equate the numerical coefficients of the terms with iden\-ti\-cal
power exponents of the variable $\mu$.\,\,We obtain a chain of
coupled linear algebraic equations for the matrices
$\mathbf{\hat{Y}}_{2}$, $\mathbf{\hat{Y}}_{3}$, and so on.\,\,But,
for the matrix $\mathbf{\hat{Y}}_{1},$ we have the quadratic
equation
\begin{equation}
\mathbf{\hat{Y}}_{1}^{2}-\mathbf{\hat{B}}_{1}\mathbf{\hat{Y}}_{1}%
-\mathbf{\hat{A}}_{1}=0, \label{sq-mat-eq}%
\end{equation}
where $\mathbf{\hat{A}}_{1}$ and $\mathbf{\hat{B}}_{1}$ are some
numerical matrices.\,\,In the case where $N=2$, the transformation
matrices can be selected to provide
$\mathbf{\hat{B}}_{1}=B_{1}\mathbf{\hat{1}}$, and the quadratic
matrix equation can be solved.\,\,Of two roots, the root
corresponding to a more regular behavior of the trial functions in
vicinities of special points must be selected (namely, the positive
root in a vicinity of $\mu=0$ and the negative one in a vicinity of
$\mu\rightarrow\infty$).\,\,The power exponents $\lambda
_{0/\infty}$ are so selected to provide the coincidence of the power
exponents in the most important terms on both sides of the equation.

In this way, the initial conditions for the matrix $\mathbf{\hat{Y}%
}_{\mathbf{S}}$ at zero and infinity were determined.\,\,The
corresponding matrices are presented in Appendix~B.\,\,Afterward,
the Riccati equations were integrated numerically, by using the
Runge--Kutta method and taking the initial conditions in the regions
$\mu\in\lbrack0,1]$ and $\mu
\in\lbrack1,\infty)$ into account.\,\,Then the values of matrices $\mathbf{\hat{Y}%
}_{\mathbf{S}}^{0/\infty}(\mu=1)$ were calculated, and the inverse
transformations were carried out to find the matrices $\mathbf{\hat{Y}%
}^{0/\infty}(\mu=1)$.\,\,A smooth matching of trial functions in the
intervals $\mu>1$ and $\mu<1$ can be done if
\begin{equation}
\delta(mR_{\mathrm{cr}},\zeta)\equiv\det[\mathbf{\hat{Y}}^{0}(\mu
=1)-\mathbf{\hat{Y}}^{\infty}(\mu=1)]=0.\,\,\label{condition}%
\end{equation}

In this work, the case $N=2$ was considered.\,\,The corresponding
transformation matrices and the matrices of initial conditions are
given in Appendix~B.\,\,All the actions described above were
executed for various values of product $mR$ and at a fixed value of
$\zeta$.\,\,A value of $mR_{\mathrm{cr}}$, at which condition
(\ref{condition}) was satisfied, was considered to be
critical.\,\,The results of numerical calculations were used to plot
the dependence $mR_{\mathrm{cr}}(\zeta)$ (solid curve in Fig.~2).

\section{Conclusions}

\label{concl}

A research of the supercritical instability in a system of two
identical charged impurities in graphene is carried out.\,\,The
charge of each impurity is selected to be subcritical, but their sum
$\zeta=2Z\alpha/\kappa$ exceeds the critical value
$\zeta_{\mathrm{c}}=1/2$.\,\,Hence, for every fixed $\zeta$, there
exist a distance $R_{\mathrm{cr}}$ between the impurities, at which
a supercritical regime is realized.\,\,The urgency of the presented
research is associated with recent observations of the supercritical
instability in clusters of Ca impurities located on graphene
\cite{Wang}.

The characteristic feature of this problem consists in that the
variables in the Dirac equation can\-not be separated in any known
coordinate system.\,\,There\-fore, it is impossible to obtain the
solution in the analytical form, and the Kantorovich varia\-tional
method was applied to find the dependence of the critical distance
on the charge of the system.\,\,For massless particles, the critical
phenomena are associated only with the emergence of a resonance in
the lower continuum.\,\,Since it is inconvenient to work with
resonances within the variational method, a gap $\Delta$ was
supposed to exist in the band spectrum of graphene.\,\,Such a gap
can be created experimentally by plenty of techniques, e.g.,
changing to a graphene nanoribbon, creating deformations,
hydrogenating the surface, and so forth \cite{Katsn}.\,\,The
presence of a gap results in the appearance of levels belonging to a
discrete spectrum.\,\,Therefore, a diving of the lowest level into
the lower continuum is observed.\,\,The distance between the
impurities, at which this diving takes place, is called critical.

The Kantorovich variational method provides an opportunity for an
arbitrary number of trial functions to be used.\,\,In this work, a
variant of the method with two trial functions was applied.\,\,For
the sake of comparison, the results of similar calculations but with
only one trial function, which were carried out in work
\cite{two-centers}, are also presented.\,\,The calculated
dependences of $mR_{\mathrm{cr}}$ on $\zeta$, as well as the
corresponding approximate curve obtained by matching the exact
solution asymptotics, are plotted in Fig.~2.\,\,One can see that the
results of two consecutive approximations by the variational method
agree rather well with each other both qualitatively and
quantitatively.\,\,The maximum discrepancy between them does not
exceed $8\div10\%$.\,\,This fact testifies that, despite the
simplicity of the applied approximation, the results obtained in
work \cite{two-centers} are satisfactory: as it should be,
$R_{\mathrm{cr}}\rightarrow0$ as $\zeta \rightarrow1/2$ (i.e.\,\,the
supercritical instability occurs only if the impurities are brought
together) and $R_{\mathrm{cr}}\rightarrow\infty$ as
$\zeta\rightarrow1$ (i.e.\,\,each impurity becomes supercritical).
It is also
demonstrated that, as $\Delta\rightarrow0$, the value $R_{\mathrm{cr}%
}\rightarrow\infty$ for any fixed $\zeta$.\,\,This means that, if
the total charge in the system exceeds the critical one, the system
is in the supercritical state at any finite distance between the
impurities.

As was shown in work \cite{two-centers}, this supercritical
instability manifests itself as the emergence of a quasi-stationary
state in the lower continuum.\,\,It can be detected in the local
density of states, which is an experimentally measured quantity.
However, the energy and the width of the resonance decrease
according to the $\frac{1}{R}$-law, as the distance $R$ between the
impurities grows.\,\,Therefore, when the distance becomes large
enough, the resonance becomes unobservable (e.g., because of a
confined accuracy of measurements).\,\,In this case, the system
state cannot be distinguished from the subcritical one.

\vskip3mm {\it The author expresses the sincere gratitude to
E.V.\,Gor\-bar and V.P.\,Gusynin for the valuable advice and
corrections made while discussing this work.\,\,The work was
sponsored by the State Fund for Fundamental Researches of Ukraine
(grant F53.2/028).}

\appendix

\subsubsection*{\!\!\!\!\!\!APPENDIX A\\
Expressions for Matrix Coefficients}

\label{A}

{\footnotesize In this Appendix, expressions for the matrices
$\mathbf{\hat{P}}$, $\mathbf{\hat{Q}}$, and $\mathbf{\hat{R}}$ in
functional (\ref{functional2}) are presented.\,\,At $E=-\Delta$,
functional (\ref{functionalS-modified}) takes the form
\[
S[\psi]=4\sum\limits_{k,l=1}^{N}\int\limits_0^\infty d\mu d\nu|J|
\biggl[(\bm\nabla\mu)^2\psi_{k}'{\psi^*_l}'\nu^{k+l-2}\,+
\]\vspace*{-3mm}
\[
+\,2\bm\nabla\mu\bm\nabla\nu\Re e(\psi^*_l\psi_k')(l-1)\nu^{k+l-3}-
\]
\[
-\,2\left(\!\frac{\bm\nabla V}{2V}\times\bm\nabla\mu\!\right)\Im
(\psi^*_l\psi_k')\nu^{k+l-2}\,+
\]
\[
+\,\psi^*_l\psi_k\biggl[(\bm\nabla\nu)^2(l-1)(k-1)\nu^{k+l-4}\,-
\]
\[
-\,i(l-k)\left(\!\frac{\bm\nabla
V}{2V}\times\bm\nabla\nu\!\right)\nu^{k+l-3}\,+
\]
\begin{equation}
+\,2U\nu^{k+l-2}\biggr]\biggr]f(\mu,\nu),
\label{functional-1}\tag{A1}
\end{equation}
where the functions $\bm\nabla\mu$, $\bm\nabla\nu$, $V$, and $U$ can
be expressed in terms of the variables $\mu$ and $\nu$.\,\,Since
$\mu\nu=\eta^{2}<1$ and $\mu(\nu+1)=\xi^{2}>1$, the integration in
the plane $(\mu,\nu)$ is carried out over the curvilinear triangle
\begin{equation}
\left(\!\frac{1}{\mu}-1\!\right)\theta\left(1-\mu\right)<\nu<\frac{1}{\mu},\tag{A2}
\end{equation}
which can be provided with the help of the function
\[
f(\mu,\nu)=\theta(1-\mu\nu)\,\times
\]
\begin{equation}
\times\,[\theta(1-\mu)\theta(\mu(\nu+1)-1)+\theta(\mu-1)].\tag{A3}
\end{equation}
Being expressed in terms of the variables $\mu$ and $\nu$, all the
quantities in functional (\ref{functional-1}) look like as follows:
\begin{equation}
|J|=\frac{\mu
R^2}{16}\frac{1}{\sqrt{\nu(\nu+1)(\mu+\mu\nu-1)(1-\mu\nu)}},\tag{A4}
\end{equation}
\begin{equation}
V(\mu,\nu)=-\frac{2v_F\zeta}{R}\sqrt{\frac{\nu+1}{\mu}},\tag{A5}
\end{equation}
\begin{equation}
(\bm\nabla\mu)^2=\frac{16}{R^2}(\mu+2\mu\nu-1),\tag{A6}
\end{equation}
\begin{equation}
(\bm\nabla\nu)^2=\frac{16\nu(\nu+1)}{\mu^2 R^2},\tag{A7}
\end{equation}
\begin{equation}
\bm\nabla\mu\bm\nabla\nu=-\frac{16\nu(\nu+1)}{R^2},\tag{A8}
\end{equation}
\[
\frac{\bm\nabla
V}{V}\times\bm\nabla\mu=(\bm\nabla\nu\times\bm\nabla\mu)
\frac{\partial\ln V}{\partial\nu}\,=
\]
\begin{equation}
=\,\frac{8}{R^2\mu}\frac{\sqrt{\nu(\mu+\mu\nu-1)(1-\mu\nu)}}{\sqrt{\nu+1}},\tag{A9}
\end{equation}
\[
\frac{\bm\nabla
V}{V}\times\bm\nabla\nu=(\bm\nabla\mu\times\bm\nabla\nu)
\frac{\partial\ln V}{\partial\mu}\,=
\]
\begin{equation}
=\,\frac{8}{R^2\mu^2}\sqrt{\nu(\nu+1)(\mu+\mu\nu-1)(1-\mu\nu)},\tag{A10}
\end{equation}
\[
2U=\frac{2}{R^2}\biggl[2v_F\zeta
mR\sqrt{\frac{\nu+1}{\mu}}-2\zeta^2\frac{\nu+1}{\mu}\,-
\]
\begin{equation}
-\,\frac{4\nu+1}{\mu}+\frac{3}{2}\frac{4\mu\nu^2+5\mu\nu+\mu-1}{\mu^2(\nu+1)}\biggr].\tag{A11}
\end{equation}
Functional (\ref{functional-1}) acquires form (\ref{functional2}), in which
$\mathbf{\hat{P}}$, $\mathbf{\hat{Q}}$, and $\mathbf{\hat{R}}$ are $N\times
N$-matrices depending on $\mu$:
\begin{equation}
\label{P} P_{kl}(\mu)=\int\limits_0^\infty
(\bm\nabla\mu)^2\nu^{k+l-2} |J|f(\mu,\nu) d\nu,\tag{A12}
\end{equation}
\[
Q_{kl}(\mu)=\int\limits_0^\infty\Biggl[(\bm\nabla\nu)^2(l-1)(k-1)\nu^{k+l-4}\,-
\]
\[
-\,i(l-k)\left(\!\frac{\bm\nabla
V}{2V}\times\bm\nabla\nu\!\right)\nu^{k+l-3}\,+
\]
\begin{equation}\label{Q}
+\,2U\nu^{k+l-2}\Biggr]|J|f(\mu,\nu) d\nu,\tag{A13}
\end{equation}
\[
R_{kl}(\mu)=\int\limits_0^\infty\Biggl[\bm\nabla\mu\bm\nabla\nu(l-1)\nu^{k+l-3}\,+
\]
\begin{equation}\label{R}
+\,i\left(\!\frac{\bm\nabla V}{2V}\times
\bm\nabla\mu\!\right)\nu^{k+l-2}\Biggr]|J|f(\mu,\nu) d\nu.\tag{A14}
\end{equation}

Let us obtain expressions for the elements of the matrices
$\mathbf{\hat{P}}$, $\mathbf{\hat{Q}}$, and $\mathbf{\hat{R}}$ in
the case $N=2$.\,\,They can be calculated with the help of integrals
taken from \cite{Gradshtein}, which are expressed in terms of the
complete elliptic integrals of the first, $K(k)$, second, $E(k)$,
and third, $\Pi(k,l)$, kinds.\,\,The final forms of the expressions
are obtained with the help of the identities \cite{Byrd}
\begin{equation}
\Pi(\mu,\mu)=\frac{\pi}{4(1-\mu)}+\frac{1}{2}K(\mu),\tag{A15}
\end{equation}
\begin{equation}
\Pi(\mu^2,\mu)=\frac{1}{1-\mu^2}E(\mu).\tag{A16}
\end{equation}
Again, since the ground state of the system is considered, and the
wave function of the ground state is real-valued, all imaginary
parts of coefficients do not make contributions.\,\,Therefore,
\begin{equation}
P_{11}(\mu)=\pi\mu,\tag{A17}
\end{equation}
\[
P_{12}(\mu)=P_{21}(\mu)=\frac{\pi}{2}(1-\mu)\,+
\]
\[
+\,\theta(1-\mu)\left[2E(\mu)-(1-\mu^2)K(\mu)\right]+
\]
\begin{equation}
+\,\theta(\mu-1)\mu
\left[2E\left(\frac{1}{\mu}\!\right)-\left(\!1-\frac{1}{\mu^2}\!\right)K\left(\!\frac{1}{\mu}\!\right)\!\right]\!,\tag{A18}
\end{equation}
\[
P_{22}(\mu)=\frac{\pi}{2}\frac{1-\mu+\mu^2}{\mu}\,+
\]
\[
+\,\theta(1-\mu)\frac{1-\mu}{\mu}\left[2E(\mu)-(1-\mu^2)K(\mu)\right]+
\]
\begin{equation}
+\,\theta(\mu-1)(1-\mu)\left[2E\left(\!\frac{1}{\mu}\!\right)-\left(\!1-\frac{1}{\mu^2}\!\right)
K\left(\!\frac{1}{\mu}\!\right)\!\right]\!,\tag{A19}
\end{equation}
\begin{equation}
R_{11}(\mu)=R_{21}(\mu)=0,\tag{20}
\end{equation}
\[
R_{12}(\mu)=-\frac{\pi}{2\mu}-\theta(1-\mu)\frac{E(\mu)}{\mu}\,-
\]
\begin{equation}
-\,\theta(\mu-1)\left[E\left(\!\frac{1}{\mu}\!\right)-\left(\!1-\frac{1}{\mu^2}\!\right)
K\left(\!\frac{1}{\mu}\!\right)\!\right]\!,\tag{A21}
\end{equation}
\[
R_{22}(\mu)=-\frac{\pi(2-\mu)}{4\mu^2}\,-
\]
\[
-\,\theta(1-\mu)\left[\frac{3-\mu}{2\mu^2}E(\mu)-\frac{1-\mu^2}{2\mu^2}K(\mu)\right]-
\]
\[
-\,\theta(\mu-1)\biggl[\frac{3-\mu}{2\mu}E\left(\!\frac{1}{\mu}\!\right)+
\]
\begin{equation}
+\,\frac{(\mu-1)(\mu^2-\mu-2)}{2\mu^3}K\left(\!\frac{1}{\mu}\!\right)\!\biggr],\tag{A22}
\end{equation}
\[
 Q_{11}(\mu)=-\frac{\pi(\zeta^2-1)}{8\mu}\,+
\]
\[
+\,\frac{\zeta mR}{2}\left[\theta(1-\mu) K(\sqrt{\mu})+
\frac{\theta(\mu-1)}{\sqrt{\mu}}K\left(\!\frac{1}{\sqrt{\mu}}\!\right)\right]+
\]
\[
+\,\theta(1-\mu)\biggl[\frac{3}{8\mu(1+\mu)}E(\mu)-
\]
\[
-\frac{(2\zeta^2+1)(1+\mu)}{8\mu}K(\mu)\biggr]+
\]
\[
+\,\theta(\mu-1)\biggl[\frac{3}{8(1+\mu)}E\left(\!\frac{1}{\mu}\!\right)-
\]
\begin{equation}\label{Q-function}
-\,\frac{(\zeta^2-1)(1+\mu)+3\mu}{4\mu^2}K\left(\!\frac{1}{\mu}\!\right)\!\biggr],\tag{A23}
\end{equation}
\[
Q_{12}(\mu)=Q_{21}(\mu)=-\frac{\pi\left(3\mu+4(\zeta^2-1)\right)}{32\mu^2}\,+
\]
\[
+\,\theta(1-\mu)\frac{\zeta mR}{2\mu}E(\sqrt{\mu})\,+
\]
\[
+\,\theta(\mu-1)\frac{\zeta mR}{2\sqrt{\mu}}\left[
E\left(\!\frac{1}{\sqrt{\mu}}\!\right)
-\frac{\mu-1}{\mu}K\left(\!\frac{1}{\sqrt{\mu}}\!\right)\right]-
\]
\[
-\,\theta(1-\mu)\left[\frac{3\mu+2(\mu+1)(\zeta^2-1)}{8\mu^2(1+\mu)}E(\mu)
+\frac{3(1-\mu)}{16\mu}K(\mu)\right]-
\]
\[
-\,\theta(\mu-1)\biggl[\frac{3\mu+2(\mu+1)(\zeta^2-1)}{8\mu(1+\mu)}
E\left(\!\frac{1}{\mu}\!\right)+
\]
\begin{equation}
+\,\frac{(1-\mu)\left(9\mu+4(1+\mu)(\zeta^2-1)\right)}
{16\mu^3}K\left(\!\frac{1}{\mu}\!\right)\!\biggr],\tag{A24}
\end{equation}
\[
Q_{22}(\mu)=\frac{\pi}{32\mu^3}\left(16-(2\zeta^2-2+3\mu)(2-\mu)\right)+
\]
\[
+\,\theta(1-\mu)\frac{\zeta m
R}{6\mu}\biggl[2\left(\!\frac{2}{\mu}-1\!\right)E(\sqrt{\mu})\,-
\]
\[
-\left(\!\frac{1}{\mu}-1\!\right)K(\sqrt{\mu})\biggr]+
\]
\[
+\,\theta(\mu-1)\frac{\zeta m
R}{6\sqrt{\mu}}\biggl[2\left(\!\frac{2}{\mu}-1\!\right)
E\left(\!\frac{1}{\sqrt{\mu}}\!\right)-
\]
\[
-\left(\!1-\frac{1}{\mu}\!\right)\left(\!\frac{3}{\mu}-2\!\right)K\left(\!\frac{1}{\sqrt{\mu}}\!\right)\biggr]+
\]
\[
+\,\theta(1-\mu)\biggl[\!\frac{(1-\mu)\left(2(\zeta^2-1)(\mu+1)-3\mu(1-\mu)\right)}
{16\mu^3}K(\mu)\,+
\]
\[
+\,
\frac{3\mu^2+13\mu+16+2(\zeta^2-1)(\mu^2-2\mu-3)}{16\mu^3(1+\mu)}E(\mu)\!\biggr]+
\]
\[
+\,\theta(\mu-1)\biggl[\!\frac{(1-\mu)(3\mu^2+5\mu+8)}{8\mu^4}
K\left(\!\frac{1}{\mu}\!\right)+
\]
\[
+\,\frac{(\zeta^2-1)(1-\mu)(\mu^2-\mu-2)}{8\mu^4}
K\left(\!\frac{1}{\mu}\!\right)+
\]
\begin{equation}
+\,\frac{3\mu^2+13\mu+16+2(\zeta^2-1)(\mu^2-2\mu-3)}{16\mu^2(1+\mu)}E\left(\!\frac{1}{\mu}\!\right)\!\biggr].\tag{A25}
\end{equation}

}

\subsubsection*{\!\!\!\!\!\!APPENDIX B\\ Characteristic features of the application\\ of the variational method in the
case \boldmath$N=2$}

\label{B}

{\footnotesize In this Appendix, some details concering the
application of the Kantorovich method in the case $N=2$ are discussed.

Transformation (\ref{transf}) changes the coefficients of
Eqs.~(\ref{Riccati-eq}) and (\ref{set-of-equations}) as follows:%
\begin{equation}
\mathbf{\hat{A}}_{\mathbf{S}}=\mathbf{\hat{S}}^{-1}\mathbf{\hat{A}}\mathbf{\hat{S}}
-\mathbf{\hat{S}}^{-1}\mathbf{\hat{B}}\mathbf{\hat{S}}'-\mathbf{\hat{S}}^{-1}\mathbf{\hat{S}}'',\tag{B1}
\end{equation}
\begin{equation}
\mathbf{\hat{B}}_{\mathbf{S}}=\mathbf{\hat{S}}^{-1}\mathbf{\hat{B}}\mathbf{\hat{S}}+2\mathbf{\hat{S}}^{-1}\mathbf{\hat{S}}'',\tag{B2}
\end{equation}
\begin{equation}
\mathbf{\hat{P}}_{\mathbf{S}}=\mathbf{\hat{S}}^{T}\mathbf{\hat{P}}\mathbf{\hat{S}},\tag{B3}
\end{equation}
\begin{equation}
\mathbf{\hat{R}}_{\mathbf{S}}=\mathbf{\hat{S}}^{T}
\mathbf{\hat{R}}\mathbf{\hat{S}}+\mathbf{\hat{S}}^{T}\mathbf{\hat{P}}\mathbf{\hat{S}}',\tag{B4}
\end{equation}
\begin{equation}
\mathbf{\hat{Q}}_{\mathbf{S}}=\mathbf{\hat{S}}^{T}\mathbf{\hat{Q}}\mathbf{\hat{S}}+
\mathbf{\hat{S}}'^{T}\mathbf{\hat{R}}\mathbf{\hat{S}}+
\mathbf{\hat{S}}^{T}\mathbf{\hat{R}}^{T}\mathbf{\hat{S}}'+
\mathbf{\hat{S}}'^{T}\mathbf{\hat{P}}\mathbf{\hat{S}}'.\tag{B5}
\end{equation}

1. In the interval $0<\mu\leq1$,%
\begin{equation}
\det\mathbf{\hat{P}}(\mu)=\frac{\pi^2
\mu^2}{8}+O\left(\mu^4\right)\!,\quad \mu\to 0.\tag{B6}
\end{equation}
Therefore, the matrix $\mathbf{\hat{S}}^{(0)}$ should be so selected
that $\det\mathbf{\hat{S}}^{(0)}\sim\frac{1}{\mu}$.\,\,In this work,
we selected the matrix
\begin{equation}
\mathbf{\hat{S}}^{(0)}(\mu)=\left(\!\!
\begin{array}{cc}
\frac{\mu-1}{\mu^{3/2}} & \frac{\mu+1}{\mu^{3/2}} \\[2mm]
 \frac{1}{\sqrt{\mu}} & -\frac{1}{\sqrt{\mu}} \\
\end{array}
\!\!\right)\!, \quad
\det\mathbf{\hat{S}}^{(0)}=-\frac{2}{\mu}.\tag{B7}
\end{equation}
The power exponent is $\lambda_{0}=-1$, which corresponds to the
power-law behavior of trial functions in a vicinity of the point
$\mu=0$.\,\,The coefficients of Eq.~(\ref{sq-mat-eq}) look like
\begin{equation}
\mathbf{\hat{A}}_{1}^{(0)}=\left(\!\!
\begin{array}{cc}
\frac{3-\zeta^2}{4} & -\frac{3}{4} \\[2mm]
-\frac{1}{4} & \frac{1-\zeta^2}{4} \\
\end{array}
\!\!\right)\!,\quad
\mathbf{\hat{B}}_{1}^{(0)}\equiv\mathbf{\hat{1}}.\tag{B8}
\end{equation}
Equation (\ref{sq-mat-eq}) is reduced to
\begin{equation}
\left(\!\mathbf{\hat{Y}}_{1}^{(0)}-\frac{1}{2}\mathbf{\hat{1}}\!\right)^{\!2}
=\mathbf{\hat{A}}_{1}^{(0)}+\frac{1}{4}\mathbf{\hat{1}}.\tag{B9}
\end{equation}
Its solution is
\[\mathbf{\hat{Y}}_{1}^{(0)}=\frac{1}{2}\mathbf{\hat{1}}+\]
\[
+\left(\!\!
\begin{array}{cc}
\frac{1}{2} & -\frac{3}{2} \\[2mm]
\frac{1}{2} & \frac{1}{2} \\
\end{array}
\!\!\right) \!\left(\!\!
\begin{array}{cc}
\frac{\sqrt{1-\zeta^2}}{2} & 0 \\[2mm]
0 & \frac{\sqrt{5-\zeta^2}}{2} \\
\end{array}
\!\!\right) \!\left(\!\!
\begin{array}{cc}
\frac{1}{2} & \frac{3}{2} \\[2mm]
-\frac{1}{2} & \frac{1}{2} \\
\end{array}
\!\!\right)=
\]
\begin{equation}
=\left(\!\!
\begin{array}{cc}
\frac{4+\sqrt{1-\zeta^2}+3\sqrt{5-\zeta^2}}{8} & \frac{3\sqrt{1-\zeta^2}-3\sqrt{5-\zeta^2}}{8} \\[2mm]
\frac{\sqrt{1-\zeta^2}-\sqrt{5-\zeta^2}}{8} & \frac{4+3\sqrt{1-\zeta^2}+\sqrt{5-\zeta^2}}{8} \\
\end{array}
\!\!\right)\!.\tag{B10}
\end{equation}
The matrix $\frac{1}{\mu}\mathbf{\hat{Y}}_{1}^{(0)}$ is used as the
initial condition, while numerically integrating the Riccati
equation (\ref{Riccati-eq}) in the interval $0<\mu\leq1$.

2. In the interval $1\leq\mu<\infty$,
\begin{equation}
\det\mathbf{\hat{P}}(\mu)=\frac{\pi^2}{8}+O\left(\!\frac{1}{\mu^2}\!\right)\!,\quad
\mu\to \infty.\tag{B11}
\end{equation}

Therefore, the matrix $\mathbf{\hat{S}}^{(\infty)}$ should be
selected so that $\det\mathbf{\hat{S}}^{(\infty)}\sim1$.\,\,In this
work, we take the matrix
\begin{equation}
\mathbf{\hat{S}}^{(\infty)}(\mu)=\left(\!\!
\begin{array}{cc}
\frac{1}{\sqrt{\mu}} & -\frac{1}{\sqrt{\mu}} \\[2mm]
\sqrt{\mu} & \sqrt{\mu} \\
\end{array}
\!\!\right)\!, \quad \det\mathbf{\hat{S}}^{(\infty)}=2.\tag{B12}
\end{equation}
The power exponent is $\lambda_{\infty}=-3/4$, which corresponds to
the behavior of the trial functions $\sim\exp(-\mu^{1/4})$ as
$\mu\rightarrow\infty$.\,\,The coefficients of Eq.~(\ref{sq-mat-eq})
look like
\begin{equation}
\mathbf{\hat{A}}_{1}^{(\infty)}=\frac{\zeta m
R}{4}\mathbf{\hat{1}},\quad
\mathbf{\hat{B}}_{1}^{(\infty)}\equiv\mathbf{\hat{0}}.\tag{B13}
\end{equation}
The solution of Eq.~(\ref{sq-mat-eq}) reads
\begin{equation}
\mathbf{\hat{Y}}_{1}^{(\infty)}=-\frac{\sqrt{\zeta m
R}}{2}\mathbf{\hat{1}}.\tag{B14}
\end{equation}
The matrix $\frac{1}{\mu^{3/4}}\mathbf{\hat{Y}}_{1}^{(\infty)}$ is
used as the initial condition, while numerically integrating the Riccati
equation (\ref{Riccati-eq}) in the interval $1\leq\mu<\infty$.

}

\vspace*{-5mm}
\rezume{%
О.О.\,Соболь}{ВАРІАЦІЙНИЙ МЕТОД\\ ОБЧИСЛЕННЯ КРИТИЧНОЇ ВІДСТАНІ В
ЗАДАЧІ\\ ДВОХ КУЛОНІВСЬКИХ ЦЕНТРІВ У ГРАФЕНІ} {Досліджено
надкритичну нестабільність в системі двох заряджених домішок у
графені з квазічастинками, що в неперервній границі описуються
двовимірним рівнянням Дірака.  Розглянуто випадок, коли заряд кожної
з двох однакових домішок є субкритичним, а їх сума перевищує
критичний заряд в задачі про один кулонівський центр. Розвинено
варіаційний метод, за допомогою якого обчислюється значення
критичної відстані між домішками $R_{\rm cr}$ як функції повного
заряду системи. Встановлено, що $R_{\rm cr}$ зростає зі збільшенням
повного заряду двох домішок та зі зменшенням ширини квазічастинкової
щілини. Проведено порівняння результатів з даними попередніх
досліджень.}

\end{document}